\documentclass[longauth]{aa} 

\usepackage{graphicx}
\usepackage{txfonts}
\usepackage{multirow}
\usepackage{booktabs}
\usepackage{xcolor}
\usepackage{gensymb}
\usepackage{textcomp}

\begin{document}

   \title{Multi-band high resolution spectroscopy rules out the\\ hot Jupiter BD+20~1790b\thanks{Based on observations made with the Italian Telescopio Nazionale Galileo (TNG) operated on the island of La Palma by the Fundaci\'on Galileo Galilei of the INAF (Istituto Nazionale di Astrofisica) at the Spanish Observatorio del Roque de los Muchachos (ORM) of the IAC.} \fnmsep\thanks{This work used the Immersion Grating Infrared Spectrometer (IGRINS) that was developed under a collaboration between the University of Texas at Austin and the Korea Astronomy and Space Science Institute (KASI) with the financial support of the US National Science Foundation under grant AST-1229522, of the University of Texas at Austin, and of the Korean GMT Project of KASI.} \fnmsep\thanks{These results made use of the Discovery Channel Telescope (DCT) at Lowell Observatory. Lowell is a private, non-profit institution dedicated to astrophysical research and public appreciation of astronomy, and it operates the DCT in partnership with Boston Univ., the Univ. of Maryland, the Univ. of Toledo, Northern Arizona Univ., and Yale Univ.} \fnmsep\thanks{This paper includes data taken with the 2.7-m H. J. Smith Telescope at The McDonald Observatory of The University of Texas at Austin.} \fnmsep\thanks{Based on observations made with the REM Telescope, INAF Chile.}}
\subtitle{First data from the GIARPS Commissioning }

\author{I. Carleo\inst{1,2},
         S. Benatti\inst{2},
         A. F. Lanza\inst{3},
         R. Gratton\inst{2},
         R. Claudi\inst{2},
         S. Desidera\inst{2},
         G. N. Mace\inst{4}, 
         S. Messina\inst{3},
         N. Sanna\inst{5},\\
         E. Sissa\inst{2},
         A. Ghedina\inst{6},
         F. Ghinassi\inst{6},
         J. Guerra\inst{6},
         A. Harutyunyan\inst{6},
         G. Micela\inst{7},
         E. Molinari\inst{6,16},
         E. Oliva\inst{5},
         A. Tozzi\inst{5},\\
         C. Baffa\inst{5},
         A. Baruffolo\inst{2},
         A. Bignamini\inst{8},
         N. Buchschacher\inst{9},
         M. Cecconi\inst{6},
         R. Cosentino\inst{6},
         M. Endl,\inst{4},
         G. Falcini\inst{5},\\
         D. Fantinel\inst{2},
         L. Fini\inst{5},
         D. Fugazza\inst{10},
         A. Galli\inst{6},
         E. Giani\inst{5},
         C. Gonz{\'a}lez\inst{6},
         E. Gonz{\'a}lez-{\'A}lvarez\inst{7,11},
         M. Gonz{\'a}lez\inst{6},\\
         N. Hernandez\inst{6},
         M. Hernandez Diaz\inst{6},
         M. Iuzzolino\inst{5,12},
         K. F. Kaplan\inst{4},
         B. T. Kidder\inst{4},
         M. Lodi\inst{6},
         L. Malavolta\inst{1},\\
         J. Maldonado\inst{7},
         L. Origlia\inst{13},
         H. Perez Ventura\inst{6}, 
         A. Puglisi\inst{5},
         M. Rainer\inst{10},
         L. Riverol\inst{6},
         C. Riverol\inst{6},
         J. San Juan\inst{6},\\
         S. Scuderi\inst{3},
         U. Seemann\inst{14},
         K. R. Sokal\inst{4},
         A. Sozzetti\inst{15},
         M. Sozzi\inst{5}
}

   \institute{Dipartimento di Fisica e Astronomia "G. Galilei"-- Universt\`a degli Studi di Padova, Vicolo dell'Osservatorio 3, I-35122 Padova \\
              \email{ilaria.carleo@oapd.inaf.it}
\and INAF -- Osservatorio Astronomico di Padova, Vicolo dell'Osservatorio 5, I-35122, Padova, Italy  
\and INAF -- Osservatorio Astrofisico di Catania, Via S. Sofia 78, I-95123, Catania, Italy 
\and Department of Astronomy and McDonald Observatory -- The University of Texas at Austin, 2515 Speedway, Stop C1400, Austin, TX, 78712-1205 
\and INAF -- Osservatorio Astrofisico di Arcetri, Largo Enrico Fermi, 5, I-50125 Firenze, Italy 
\and Fundaci\'on Galileo Galilei - INAF, Rambla Jos\'e Ana Fernandez P\'erez 7, E-38712, Bre\~na Baja, TF - Spain  
\and INAF -- Osservatorio Astronomico di Palermo, Piazza del Parlamento, 1, I-90134, Palermo, Italy 
\and INAF -- Osservatorio Astronomico di Trieste, via Tiepolo 11, I-34143 Trieste, Italy  
\and Département d'Astronomie -- Université de Genève, Chemin des Maillettes, 51, CH-1290, Versoix, Switzerland 
\and INAF -- Osservatorio Astronomico di Brera, Via E. Bianchi 46, I-23807 Merate (LC), Italy  
\and Dipartimento di Fisica e Chimica - Universit\`a degli Studi di Palermo,Via Archirafi 36, 90123 Palermo, Italia 
\and Officina Stellare S.r.l., Via Della Tecnica, 87/89, I-36030 Sarcedo (VI) - Italy
\and INAF – Osservatorio Astronomico di Bologna, Via Gobetti 93/3, I-40129, Bologna, Italy
\and Institut f\"ur Astrophysik -- Georg-August-Universit\"at G\"ottingen, Friedrich-Hund-Platz 1, D-37077 G\"ottingen, Germany 
\and INAF -- Osservatorio Astrofisico di Torino, Via Osservatorio 20, I-10025, Pino Torinese (TO), Italy
\and INAF -- Osservatorio Astronomico di Cagliari, Via della Scienza 5, I-09047, Selargius (CA), Italy
}

   \date{Received ; accepted 2018 April, 17th}

 
  \abstract
   {Stellar activity is currently challenging the detection of young planets via the radial velocity (RV) technique. }
  {We attempt to definitively discriminate the nature of the RV variations for the young active K5 star BD+20~1790, for which visible (VIS) RV measurements show divergent results on the existence of a substellar companion.   }
   {We compare VIS data with high precision RVs in the near infrared (NIR) range by using the GIANO -- B and IGRINS spectrographs. In addition, we present for the first time simultaneous VIS-NIR observations obtained with GIARPS (GIANO -- B and HARPS -- N) at Telescopio Nazionale Galileo (TNG). Orbital RVs are achromatic, so the RV amplitude does not change at different wavelengths, while stellar activity induces wavelength-dependent RV variations, which are significantly reduced in the NIR range with respect to the VIS.}
   {The NIR radial velocity measurements from GIANO -- B and IGRINS show an average amplitude of about one quarter with respect to previously published VIS data, as expected when the RV jitter is due to stellar activity. Coeval multi-band photometry surprisingly shows larger amplitudes in the NIR range, explainable with a mixture of cool and hot spots in the same active region.} 
   {In this work, the claimed massive planet around BD+20~1790 is ruled out by our data. We exploited the crucial role of multi-wavelength spectroscopy when observing young active stars: thanks to facilities like GIARPS that provide simultaneous observations, this method can reach its maximum potential.}


\titlerunning{}
\authorrunning{I. Carleo et al.}
   \maketitle
%

\section{Introduction}
\label{sec:intro}
The detection of giant planets around young stars can address key questions in the astrophysics of planetary formation and migration. In fact, looking at young stars provides an opportunity to observe the architecture of planets in their infancy. A variety of physical processes, such as planet-disc interaction, the Kozai mechanism \citep{Kozai}, planet-planet scattering \citep{Baruteau}, or in-situ formation \citep{Batygin}, responsible for generating hot Jupiters (HJs), are expected to produce observable effects, including differences in orbital parameters (eccentricity and/or obliquity), migration timescales, an age-dependent frequency of such systems, and differences in the atmospheric composition of hot Jupiters across stellar ages. The investigation of the formation and migration histories may be carried out for example via intensive radial velocity (RV) monitoring of a sample of young stars, finding new planets and testing possible differences in the frequency of hot Jupiters with age. However, the high level of stellar activity characterizing young stars induces RV variations able to mimic planetary signals, demanding a specific processing in order to investigate its actual contribution. Recently, noteworthy results have been obtained through the modelling of stellar activity as correlated noise in a set of data with the Gaussian processes regression (see e.g. \citealt{haywood2014,damasso2018}).
The first pioneering detection of a young HJ with the RV technique was recently announced around the weak-line T Tauri star V830 Tau using spectropolarimetry (\citealt{Donati2016, Yu, Donati2017}), after years of attempts marred by difficulties in dealing with the identification of planetary signals (both spectroscopic and photometric) in the presence of very high levels of activity of their young hosts.
Simultaneous multi-band high resolution spectroscopy (HRS), for example in the visible (VIS) and near infrared (NIR, 700 - 2500 nm) ranges, is a powerful tool for disentangling wavelength-dependent activity-induced RV signals (the impact of activity on RVs is expected to be typically three times lower in the NIR than in the VIS; see \citealt{Prato, Mahmud, Crockett}) from those Keplerian that are achromatic.
An emblematic example of this principle is represented by the case of the HJ around TW Hya claimed by \cite{Setiawan}, which was ruled out by NIR observations \citep{Figueira2}. \\
The building of new instrumentation allowing high precision HRS in the NIR range has been mainly pushed by the RV search of potentially habitable rocky planets around M-dwarf stars (see e.g. \citealt{Reiners}, \citealt{bonfils} and references therein). Some of those instruments, such as CARMENES (520--1710 nm, with spectral resolutions R=80,000 -- 100,000, \citealt{carmenes,reiners2016}) or HARPS+NIRPS (0.95--1.8 nm, R$\sim$100,000, \citealt{nirps}), also guarantee contemporary VIS-NIR observations. 
In this framework, the GIARPS (GIANO --B + HARPS -- N) project \citep{Claudi} has been conceived precisely to exploit the full potential of the simultaneous VIS and NIR HRS technique, allowing us to have the two high resolution spectrographs of the Telescopio Nazionale Galileo (TNG, La Palma) working simultaneously: HARPS -- N (High Accuracy Radial velocity Planet Searcher for the Northern emisphere) in the visible and GIANO -- B in the NIR (see details in Sect. \ref{sec:instr}).
The RV precisions achievable by the VIS and NIR channels of GIARPS (see more details in Sect. \ref{sec:instr}) allow us to identify giant planets signals as well as to discriminate the high activity levels of the host stars, enabling an overall monitoring of stellar activity and simultaneous detection of atomic and molecular species in planetary atmospheres.

In this paper, we present the RV measurements of a debated HJ around BD+20~1790 (V429 Gem, K5Ve, V=9.9, \citealt{Jeffries}), a very active star ($\log R'_{HK}=-3.7$, \citealt{Hernan2}, hereafter HO15) and probable member of the AB Dor moving group \citep{torres08} with an adopted age of $149^{+51}_{-19}$ Myr \citep{Bell}. 
\noindent After dedicated spectroscopic observations, \citealt{Hernan1} (hereafter HO10) interpreted the RV variation of this target as being due to the presence of a massive HJ with a period of 7.8 days. \cite{Figueira} questioned the planet, providing CORALIE RVs showing a clear correlation with the bisector span (BIS), thus attributing the RV variations to photospheric processes. Later, HO15 reported a new RV and activity study exploiting a larger spectroscopic and photometric dataset supporting again the presence of a HJ around BD+20~1790. Finally, \cite{Gagne} published a small number of CSHELL NIR RVs with no conclusive results.

This paper is organized as follows: we describe the instruments used in Sect. \ref{sec:instr}, then we present the observations and the characteristics of our dataset in Sect. \ref{sec:obs}. In Sect. \ref{sec:res} we show the analysis of both the spectroscopic and photometric time series, suggesting a theoretical model to explain the observed stellar behaviour in the VIS and NIR bands in Sect. \ref{sec:disc}. Finally we draw our conclusions in Sect. \ref{sec:conc}.


\section{Instruments}
\label{sec:instr}
In this section we summarize the characteristics of the instrumentation used to collect the data of BD+20~1790, with particular emphasis on the new TNG facility GIARPS. In 2012 the high resolution spectrograph HARPS -- N \citep{cosentino2014} was installed at the Nasmyth -- B focal station of the TNG. HARPS -- N works in the visible range ($0.39-0.68$ $\mu$m) with a resolution of R $=115,000$. In the same year, the NIR high resolution spectrograph GIANO \citep{Oliva}, working in the wavelength range from $0.97$ to $2.45\ \mu$m at a resolution of R $50,000$, was installed and commissioned at the Nasmyth -- A focal station of the TNG in 2014. Although designed for direct light feed from the telescope, in its first phase GIANO was fed by fibers. Through the "WOW" (a Way to Others Worlds) Progetto Premiale funding scheme, the Italian National Institute for Astrophysics (INAF) proposed to move GIANO to the Nasmyth -- B (re-naming the instrument GIANO -- B) and to carry out its simultaneous use with HARPS -- N aiming to achieve high resolution spectroscopy in a wide wavelength range ($0.383$ - $2.45\ \mu$m with a small gap between $\sim 0.7$ and $0.950\ \mu$m) obtained in a single exposure. The focus change, made in 2016, also allowed us to restore the original direct feeding from the telescope \citep{Tozzi} and the coupling with HARPS -- N. This was the beginning of the observing mode called GIARPS (GIANO -- B \& HARPS -- N, \citealt{Claudi}). The two spectrographs are still able to work separately, so it is possible to have three different observing modes: a) HARPS -- N only; b) GIANO -- B only; c) GIARPS, by splitting the light with a dichroic. GIARPS uses both the instruments for high precision RV measurements, exploiting the simultaneous reference technique with HARPS -- N in the visible (providing a RV precision $\sigma_{\rm{RV}}$ of $ \sim 1.0$ m\, s$^{-1}$) and a Cross-Correlation Function (CCF) method based on telluric lines with GIANO -- B ($\sigma_{\rm{RV}}$ $\sim 10$ m\, s$^{-1}$, \citealt{Carleo}). The introduction of an absorption gas cell \citep{seemannetal2014}, foreseen at the end of 2018, will allow us to reach a better precision ($\sim$ 3 m\,s$^{-1}$) in the NIR, since the spectral features of a gas cell are more reliable in comparison with the instability of the telluric spectrum. Because of its characteristics, GIARPS can be considered the first and unique worldwide instrument providing not only high resolution but also high precision radial velocity measurements in such a large wavelength range between B and K bands.\\

IGRINS (Immersion Grating Infrared Spectrometer) \citep{mace} is the cross-dispersed NIR spectrograph, mounted alternatively at the Harlan J. Smith 2.7 m telescope (McDonald Observatory, TX, USA) and at the 4.3 m Discovery Channel Telescope (Lowell Observatory, AZ, USA), with a resolving power of R=$45,000$ \citep{Yuk,Park}. It covers the H and K windows, from 1.45 to 2.5 $\mu m$ in a single acquisition. REM (Rapid Eye Mount; \citealt{Chincarini}) is a 60 cm robotic telescope located at the La Silla station of the European Southern Observatory (ESO, Chile). 
The telescope hosts two instruments: REMIR, an infrared imaging camera, and ROS2, a visible imager. The two cameras can also observe simultaneously the same field of view thanks to a dichroic placed in front of  the telescope focus. 
The ROS2 camera is equipped with a back-illuminated CCD (2048$\times$2048 pixels, 13.5 micron pixels size, 0.58 arcsec$/$pixel plate scale), which has a corrected 9.8$\times$9.8  arcmin field of view, and observes simultaneously through the four Sloan$/$SDSS g,r,i,z filters. The REMIR camera is equipped with a Hawaii I CCD (512$\times$512 pixels, 1.2 arcsec$/$pixel plate scale), which has a corrected 10$\times$10 arcmin field of view, and it is equipped with zJHK filters.


\section{Observations and data reduction}
\label{sec:obs}
In this section we describe the data acquired from each instrument. The spectroscopic datasets are presented in Sects. \ref{sec:gianodata},  \ref{sec:harpsndata}, and \ref{sec:igrinsdata}, and a complete summary of them, including the uncertainties of the RVs measurements (obtained as explained before), the typical signal-to-noise ratio (S/N), the RV r.m.s., and the peak-to-valley amplitude, is reported in Table \ref{tab:targets}. The photometric observations are described in Sec. \ref{sec:REMdata}. \\
\subsection{\textit{GIANO/GIANO -- B}}
\label{sec:gianodata}
We collected spectra of BD+20~1790 with TNG instrumentation in three different observing campaigns (see Table \ref{tab:targets}). 
\begin{table*}[htbp]
\centering
\caption{\label{tab:targets} Summary of the spectroscopic data presented in this work. For each dataset we list the instrument used for the observations, the number of spectra, the spectral range, the typical SNR, the RV nominal internal error ($\sigma_{RV}$), the RV r.m.s. scatter, and the peak-to-valley value of the RVs.}

\begin{tabular}{lcccccc}
\hline
\noalign{\smallskip}
Instrument    &  N$_{\rm spectra}$   &  Spect. range   & SNR     &   $\sigma_{RV}$     &    RV r.m.s.     &   Peak-to-valley  \\
   &  & ${\rm (\mu\rm m)}$  &            &  (kms$^{-1}$)       &  (kms$^{-1}$)    &  (kms$^{-1}$) \\
\hline  
\noalign{\smallskip}
GIANO/GIANO -- B &   18  &  0.95 - 2.45  &        72      &       0.036          &   0.130      &       0.384\\
\noalign{\smallskip}
\hline
\noalign{\smallskip}
HARPS -- N   &  20  &  0.38 - 0.69  &    35     &       0.029      &    0.280    & 1.036 \\
\noalign{\smallskip}
\hline
\noalign{\smallskip}
IGRINS  &  29  & 1.45 - 2.5 &   134     &       0.059          &       0.109      &       0.384\\
\noalign{\smallskip}
\hline
\end{tabular}
\end{table*}
The first dataset consists of 18 observations with GIANO (March 28 -- April 3, 2016), supported by quasi-simultaneous photometry with the REM telescope (see below).
The second dataset (seven spectra) was acquired during the commissioning of GIANO -- B in November 2016. The third one contains five spectra of GIANO -- B obtained during the GIARPS commissioning in March 2017. Two of them are acquired in GIARPS mode in order to test the simultaneity of GIANO -- B and HARPS -- N observations. One additional HARPS -- N spectrum was later collected without the NIR counterpart because GIANO -- B was temporarily unavailable.

GIANO data were reduced with the IRAF package ECHELLE and the dedicated scripts collected in the GIANO\_TOOLS\footnote{Available at the TNG webpage: \url{ http://www.tng.iac.es/instruments/giano/giano_tools_v1.2.0.tar.gz}.} package, while GIANO -- B spectra are processed with the dedicated pipeline GOFIO (Rainer et al, SPIE, in prep).
We obtained a set of 30 NIR RVs over one year (listed in Table \ref{tab:bd201790_giano}), with the method described in \cite{Carleo}, in which the telluric lines are used as wavelength reference and the CCF method is used to determine the stellar RV. For this purpose, we constructed two suitable digital masks that include about 2000 stellar lines and a similar number of telluric lines. After the correction of the spectra to the barycentre of the solar system, the procedure performs the cross correlation of individual orders of the normalized spectra with the appropriate masks (both stellar and telluric), with the derivation of individual CCF. Through a weighted sum of the CCFs of the individual orders, we obtain the final stellar and telluric CCFs. These are fitted with Gaussian profiles to derive the stellar and telluric RV, respectively. The latter are finally subtracted from the former, providing the relative stellar RVs. The uncertainties are then evaluated taking into account the photon statistics. 
In the present analysis we consider a slightly different approach, which takes into account the weight of the single orders, some of them being affected by telluric lines and thus contributing in different ways to the determination of the RVs. Starting from the RV of individual orders, we calculate the weighted average RV for each exposure and its corresponding error. As a final step, we derive the bisector velocity span (BIS, as in \citealt{Carleo}) of the CCF and we calculate the uncertainties on this quantity by considering the fractions of the CCF used for the derivation of the BIS, resulting in $\sqrt{10}{\sigma_{RV}}$, where $\sigma_{RV}$ is the RV error.

\subsection{\textit{HARPS -- N}}
\label{sec:harpsndata}
HARPS -- N RVs are extracted with the usual data reduction software (DRS, \citealt{Pepe02}) by cross-correlating the observed spectrum with a numerical mask that depicts the spectral features of a K5 star. To work with uniform RV values, we processed our HARPS -- N spectra and the ones collected by  HO15 (except for one spectrum with very low signal-to-noise ratio at JD 2456681) with the current HARPS -- N DRS through the YABI workflow \citep{yabi} installed at IA2\footnote{http://ia2.inaf.it.} at the INAF Observatory of Trieste. Since this facility allows us to customize the re-processing, we enlarge the width of the CCF to take into account the quite large $v\sin i$ of this star (10.03$\pm$0.47 km\, s$^{-1}$, \citealt{lopez}) and the consequent line broadening. The resulting RVs are listed in Table \ref{tab:bd201790_harpsn}, together with the CaII activity indicator, $\log$R'$_{HK}$, obtained with the dedicated tool of the HARPS -- N DRS  (the method is provided in \citealt{lovis}) also available on YABI, and the BIS (see e.g. \citealt{Queloz2001}) estimated as in Lanza et al. (submitted), starting from the computed CCF of the DRS. The usually adopted uncertainty for the BIS is twice the value of the RV uncertainty, on the basis that the BIS slope is calculated using the top and lower half of a single line measurement.
However, this occurs in the ideal case in which the bins of the CCF used for the estimation of the bisector are not correlated, therefore, in our case the multiplying factor is increased to 2.5 to avoid an underestimation of the uncertainties (see the justification in Lanza et al. (accepted)).

\subsection{\textit{IGRINS}}
\label{sec:igrinsdata}
The 29 spectra of BD+20~1790 collected with IGRINS over one year, from April 2016 to March 2017, are reduced with the IGRINS Pipeline Package.\footnote{\url{https://github.com/igrins/plp}.}
We acquired three different datasets, since the run with IGRINS installed at the Discovery Channel Telescope (DCT) occurred in between the two runs at the McDonald (McD) Observatory (see Table \ref{tab:targets}). The RV measurements (Table \ref{tab:bd201790_igrins}) are obtained with the same procedure used for GIANO.\\

\subsection{\textit{REM}}
\label{sec:REMdata}
During the first run with GIANO we obtained quasi-simultaneous photometry with REM from May 30 to April 4, 2016. We used tasks within IRAF for bias correction and flat fielding, and the technique of aperture photometry to extract magnitude time series for BD+20~1790 and for other stars detected in the frames, which were selected as candidate comparison stars.
In particular we identified two stars that were found to be non-variable and were used as comparison (C) and check (CK) stars (C: 2MASS 07233899+2025102, J = 11.23 mag, H = 11.43 mag, K = 11.17 mag; CK: 2MASS 07234597+2025328,  J = 12.15 mag, H = 11.97 mag, K = 11.91 mag). We measured a standard deviation $\sigma_{C-CK}=$ 0.014 mag in their differential light curve. The visible and infrared magnitudes of BD+20~1790 were computed differentially with respect to the comparison star.
After averaging the six consecutive differential magnitudes obtained on each night with ROS2, we obtained a time series of six average griz-band differential magnitudes and four JHK-band differential magnitudes for the subsequent analysis. The average standard deviation associated with the nightly averaged magnitudes, which we consider as our photometric precision, is $\sigma <$ 0.015 mag for all filters, and $\sigma \simeq$ 0.04 mag for the H and K magnitudes.


\section{Data analysis}
\label{sec:res}


\subsection{Spectroscopic data}
\label{subs:spec}
\begin{figure}[htbp]
    \centering
    \includegraphics[width=0.60\linewidth,angle=270]{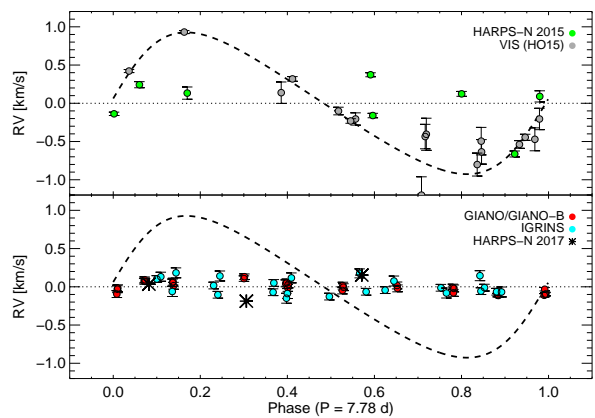}
    \caption{Orbital fit at 7.78 days found by HO15 compared to phase-folded visible and NIR  RVs. Top panel: Orbital fit (black dashed line) obtained with the visible data (FOCES, SARG, and HERMES RVs from HO15, grey dots) and HARPS-N 2015 RVs (green dots). Bottom panel: Orbital fit (black dashed line), GIANO/GIANO-B (red dots), IGRINS (light blue dots), and HARPS-N 2017 (black asterisks, two acquired in GIARPS mode) RVs.}
    
    \label{fig:GIANO_orb}
    \vspace{-0.3cm}
\end{figure}

First, we reproduced the orbital fit of the VIS RVs as presented in HO15 (with SARG, FOCES, and HERMES data only) with a Keplerian function\footnote{ We adopted the IDL least-squares MPFIT package available at \url{http://purl.com/net/mpfit}.} as in \cite{desidera11}. Our model is displayed in the upper panel of Fig. \ref{fig:GIANO_orb} as a dashed line together with the VIS data used for the fit, represented by grey dots.
The obtained RV semi-amplitude is 926.8 $\pm$ 34.6 ms$^{-1}$ and the period is 7.7827 $\pm$ 10$^{-4}$ days, in good agreement with the result proposed by HO15.
As in HO15, we were not able to obtain a solution including their HARPS -- N data so we only over-imposed them to the fitting function. We observe a match that by itself casts some doubt on a Keplerian interpretation of the RV variation.

Our phase-folded NIR RVs are shown as well in Fig. \ref{fig:GIANO_orb} (lower panel): GIANO and GIANO -- B RVs (red dots) show an r.m.s. scatter of 129.6 ms$^{-1}$, while IGRINS RVs (light blue dots) show an r.m.s. scatter of 109.2 ms$^{-1}$. The r.m.s. scatter of the whole NIR dataset is 119 ms$^{-1}$. Finally, the black asterisks represent new HARPS -- N data (two of them have the NIR simultaneous counterpart). Figure \ref{fig:GIANO_orb} shows that the amplitude of NIR RVs  (calculated as the difference between maximum and minimum RVs) is 437.3 ms$^{-1}$, four times lower with respect to the optical one reported in HO15 and interpreted as a signature of a hot Jupiter. Therefore, according to our data we can exclude any companion with those characteristics, ascribing the observed variation to phenomena of stellar origin.
\begin{table}
\caption[]{Summary of the values of Spearman ($\rho$) correlation coefficients and corresponding $p$-values between RVs and activity indicators.}
\label{corr}
$$
\begin{array}{p{0.5\linewidth}lcc}
\hline
\noalign{\smallskip}
Parameters & \rho & p\mbox{-value}\\
\hline
\noalign{\smallskip}
HARPS -- N (2015, 2017): & & \\
RV - BIS & -0.92 & 2 \cdot 10^{-8}  \\
RV residuals - BIS  & 0.39 & 0.13  \\
RV - $\log R'_{HK}$ & 0.23 & 0.45 \\
RV - v$_{\rm{asy(mod)}}$ & 0.89 & 2 \cdot 10^{-7} \\
RV - $\Delta V$ & -0.91 & 1 \cdot 10^{-7} \\
RV - H${\alpha}$ & -0.35 & 0.22  \\
\hline
\noalign{\smallskip}
GIANO,GIANO -- B: & & \\
RV - BIS &  -0.14 & 0.44   \\
RV - HeI &   -0.08 & 0.65 \\
\hline
\noalign{\smallskip}
IGRINS: & & \\
RV - BIS & 0.13 &  0.48   \\
RV - Br$\gamma$ & 0.12 & 0.51 \\
\noalign{\smallskip}
\hline
\noalign{\smallskip}
\end{array}
$$
\end{table}

\begin{figure}[htbp]
    \centering
    \includegraphics[width=0.57\linewidth, angle=270]{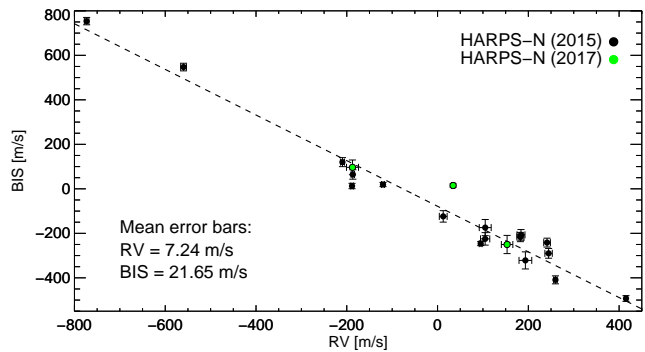}
    \caption{Correlation between RVs and BIS for HARPS -- N data (HO15 and the three new epochs in this work).}
    \label{fig:RV-BIS}
    \vspace{-0.4cm}
\end{figure}

\begin{figure}[htbp]
    \centering
    \includegraphics[width=0.55\linewidth,angle=270]{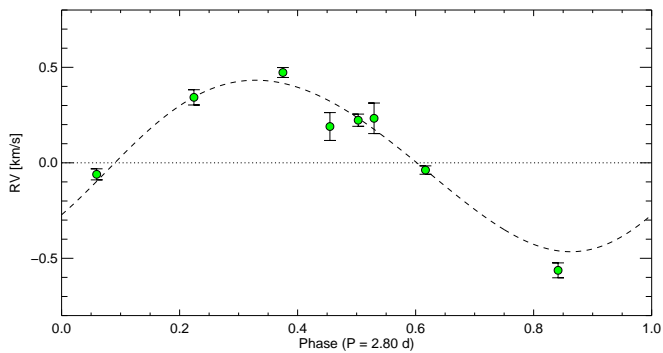}
    \caption{Phase-folded HARPS -- N RVs (2015, reprocessed from HO15 dataset) at stellar rotational period. 
    }
    \label{fig:VIS_rot}
    \vspace{-0.2cm}
\end{figure}

Actually, the Spearman rank correlation  between RVs and BIS, in this case for the whole HARPS -- N dataset (HO15 and the three epochs presented here) is -0.92, with a very high statistical significance (p-value = $2 \cdot 10^{-8}$, evaluated through the IDL Astronomy Library routine {\tt SAFE\_CORRELATE}; see Table \ref{corr} for a summary of the measured correlations between RVs and activity indices of the whole dataset), showing an unambiguous linear trend (Fig. \ref{fig:RV-BIS}). 
After subtracting this correlation from the HARPS -- N RVs time series, the resulting residuals show an r.m.s. of 61.1 ms$^{-1}$ and the Generalized Lomb-Scargle (GLS) periodogram \citep{Zechmeister} does not show any significant periodicity (i.e. with an amplitude larger than four times the standard deviation of the power spectrum, corresponding to a false alarm probability (FAP) larger than 0.01). As further evidence that the original HARPS -- N RVs are modulated by the stellar rotation of 2.801 $\pm$ 0.001 d (the photometric period by HO10), in Fig. \ref{fig:VIS_rot} we fit them with a Keplerian function by using as a first guess that period, obtaining a very good agreement with our model (P=2.80150$\pm$2$\cdot 10^{-5}$ d, K=0.449$\pm$7.370 kms$^{-1}$, e=0.065$\pm$0.018). 

We also investigated the other activity indices available from HARPS -- N, for example $\log R'_{HK}$ as described in Sect \ref{sec:obs}, and the CCF asymmetry indices: the $\Delta $V (a measure of the RV shift
produced by the exclusive contribution of the asymmetry of the CCF; see \citealt{nardetto} and \citealt{figueira2013}) and the V$_{\rm{asy(mod)}}$ (the modified version of the index V$_{\rm{asy}}$ by \citealt{figueira2013} for which the dependence from the RV shift is removed), presented in Lanza et al. (submitted).\footnote{See a description of the V$_{\rm{asy(mod)}}$ in the poster ``Line asymmetry indicators to detect stellar activity effects in radial velocity measurements" by Lanza et al., available at:  \url{https://sites.google.com/a/yale.edu/eprv-posters/home}.} Finally, we also checked the correlation with the H$\alpha$ index derived as in \cite{Sissa}. As for the BIS, the other asymmetry indices show significant correlation with RVs, while we find weak correlations with $\log R'_{HK}$ and H$_{\alpha}$ indicators (Table \ref{corr}).
We then focused our attention on the NIR data. We computed the GLS for both GIANO and IGRINS RVs even if the sampling is not suitable for a proper periodogram analysis (too few and too sparse points) or for a proper resolution of the known photometric period. The periodograms are quite noisy and do not exhibit any significant periodicity. Only for GIANO did we investigate the FAP (estimated by generating 10,000 artificial RV curves obtained from the real one, keeping the epochs of observations fixed but making random permutations of the RV values) of a peak at 7.74 d  responsible for the apparent signal of the GIANO data when phase-folded with the orbital period proposed by HO15 (red dots in Fig. \ref{fig:GIANO_orb}, lower panel). This test returned a FAP of 24.4\%, so this periodicity is probably produced by random noise. We also noticed a small amount of power corresponding with the photometric period of the star, but it appears to be related to the GIANO data sampling according to the analysis of the window function.
No prominent periodicity is found either when examining the GLS of the whole NIR dataset, which is obtained by applying a quite negligible RV offset between the two instruments (5 m s$^{-1}$).

In order to investigate a possible correlation between NIR RVs and activity indicators, we first measured the Spearman coefficient for GIANO and GIANO -- B data, in particular between the BIS and the RVs, the BIS and HeI index at 1.083 $\mu m$ (extracted as in \citealt{Robertson}), and between the HeI index and the RVs, but we found no strong correlation (Table \ref{corr}), mainly due to the uncertainties in the measurements. 
A similar analysis was performed for the IGRINS data. Since the HeI line is out of the spectral coverage of IGRINS, we investigated the impact of the activity on RVs through the Brackett $\gamma$ (Br$\gamma$) emission line at 2.16 $\mu m$ (not available in the GIANO spectra because of the discontinuity among the orders), extracted as in \cite{Robertson}. As in the case of GIANO, 
no significant correlation was found between RVs and these indicators (Table \ref{corr}). Apparently our NIR radial velocities for BD+20~1790 are not highly sensitive to activity.


\subsection{Photometric data}
\label{sec:photom}

\begin{figure}[htbp]
    \centering
    \includegraphics[width=0.80\linewidth, trim=0.2cm 1cm 1cm 2.7cm,clip]{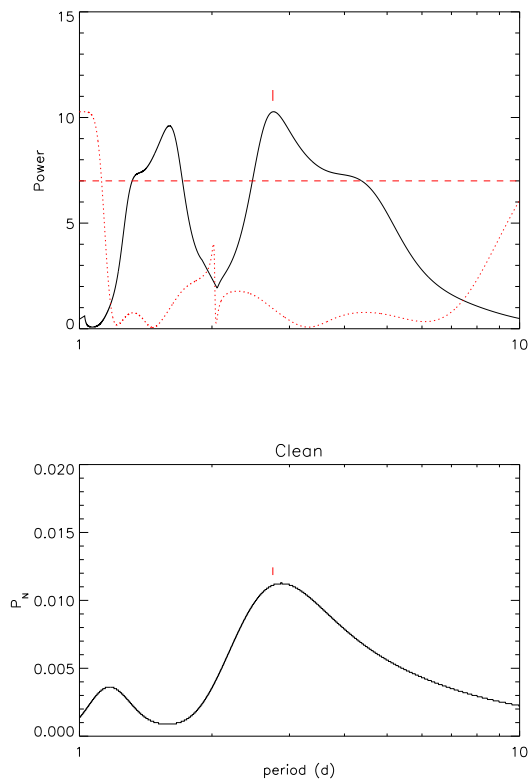}
    \caption{Top panel: Generalized Lomb-Scargle periodogram of the g-band photometric time series of BD+20~1790. The solid black line is the normalized power versus period, the dotted red line is the spectral window, and the horizontal dashed line indicates the power level corresponding to a FAP = 0.01. Bottom panel: CLEAN periodogram. The red mark is the most powerful and significant peak in the periodogram.}
    \label{fig:phot_periodogram}
    \vspace{-0.3cm}
\end{figure}

We used the GLS and the CLEAN \citep{Roberts} periodogram analyses to search for significant periodicities in the BD+20~1790 photometric time series related to its rotation period. As an example, the GLS and CLEAN periodograms are plotted in Fig. \ref{fig:phot_periodogram} for the case of the g filter. The solid black line represents the normalized power versus period, whereas the dotted red line is the spectral window function. The horizontal red dashed line represents the power level corresponding to a FAP = 0.01 (confidence level of 99$\%$, obtained with 1,000 moke light curves as in Sect. \ref{subs:spec}).  

\vspace{0.5cm}
Our periodogram analysis of the photometric variation confirmed the already known stellar rotational period P = 2.76 $\pm$ 0.04 days (corresponding to the most powerful and significant peak in the periodogram, indicated with a red mark), slightly lower than the literature value possibly because of the short baseline of the observations (six consecutive nights) or the effect of differential rotation. The uncertainty on the period is calculated following the prescription of \cite{Lamm}.

\begin{figure}[htbp]
    \centering
    \includegraphics[width=0.80\linewidth, trim=0.2cm 3cm 1cm 2.7cm,clip]{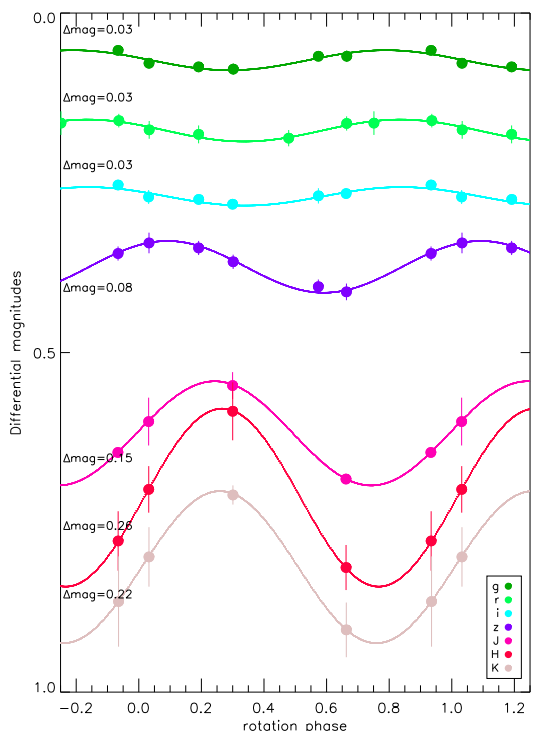}
    \caption{Differential lightcurves of BD+20 1790 phased with the rotation period P = 2.76d in different photometric bands. The solid line is a sinusoidal fit to the data with the  same period. Labels show the peak-to-peak amplitudes of the lightcurves.}
    \label{fig:phot_lightcurves}
\end{figure}

We note that the secondary power peak in the GLS periodogram is absent in the CLEAN periodogram, which has the capability of effectively removing beat frequencies arising from the data sampling.\\
In Fig. \ref{fig:phot_lightcurves} we plot the differential light curves of BD+20 1790 phased with the rotation period P = 2.76 d, using different colours for different filters. The solid lines represent sinusoidal fits to the phased magnitude computed using the rotation period. The peak-to-peak amplitudes of the sinusoidal fits measured in different filters are reported as labels in the figure. We observe that the REM light curves show different amplitudes in the different bands: unexpectedly, the amplitudes at longer wavelengths are greater than those at shorter wavelengths and the NIR modulation is almost in anti-phase with respect to the optical modulation in the $gri$ passbands (the maximum of the light curves are located at $\sim0.8$ for the $gri$ filters, at $\sim0.1$ for $z$ and at $\sim0.2$ for JHK). 
Moreover, the amplitudes of the optical light curves are a factor of approximately seven. smaller than in other seasons (cf. the light curves in Fig. 3 in HO15). This could be due to a transient phase of peculiar activity during the REM observing season (that covered a very short period) with respect to what was observed in past campaigns. This particular variability is also supported by the comparison between our HARPS -- N RVs and the ones in HO15, as noticed also by \cite{Figueira} with CORALIE data.

 \section{Discussion}
 \label{sec:disc}
The present investigation adds another important piece of evidence to the multi-wavelength characterization of the RV variations in late-type stars. Our results show that the amplitude of the NIR RV modulation in BD$+$20~1790 is generally a factor of $2-3$ smaller than in the optical band, in agreement with the results obtained by \cite{Crockett} for very active young stars. This indicates that brightness inhomogeneities, whose contrast is generally smaller in the NIR than in the optical passband, are mainly responsible for the RV variations via line profile distortions. Other effects, such as quenching of convective blueshifts (e.g. \citealt{Lanza2011}) or line profile distortions produced by the Zeeman effect \citep{Reiners2013}, are probably less important in these very active and rapidly rotating ($v\sin i \geq 10$ km~s$^{-1}$) stars. Nevertheless, an intriguing result is the small amplitude of the RV variations in the NIR as measured by GIANO contemporaneously to the REM observations that show a remarkable NIR rotational modulation in the J, H, and K passbands. To interpret these results, we first consider a simple model for the wide-band photometric variations, including the effects of both dark and bright spots. Several models for the simultaneous photometric and RV variations of late-type stars have been proposed (e.g. \citealt{Boisse2012}; \citealt{Dumusque2014}; \citealt{Herrero2016}). They include the effects of solar-like faculae whose contrast increases towards the limb. Here, we consider a hot spot in the photosphere that has a constant contrast at different limb positions, similarly to the behaviour generally assumed for a cool spot.
 
From the stellar $v\sin i$, radius, and rotation period (from HO10), we estimated an inclination of the stellar spin axis to the line of sight of $\sim 50^{\circ}$. With a simulation we then reproduced the sinusoidal shapes, amplitude ratio, and phase difference of the optical and the NIR light curves by assuming a circumpolar active region, always in view, consisting of two co-spatial components (as observed in e.g. V410 Tau, \citealt{Rice}): a cool feature (hereinafter "cool spot") covering a fraction $f_{\rm s}$ of its total area  and a hot feature  (hereinafter "hot spot") covering the remaining fraction $1-f_{\rm s}$. \\
Therefore, the average brightness of the active region at wavelength $\lambda$ can be written as 
\begin{equation}
I_{\rm a} = f_{\rm s} B(T_{\rm c}, \lambda)+ (1-f_{\rm s}) B(T_{\rm h}, \lambda),
\end{equation}
 where, for simplicity, we assume that the brightness of each component is given by a Planck function $B(T, \lambda),$ with $T_{\rm c}$ being the temperature of the cool spot and $T_{\rm h}$ that of the hot spot. Those temperatures verify the inequality: $T_{\rm c} < T_{\rm phot} < T_{\rm h}$, where $T_{\rm phot}$ is the temperature of the unperturbed photosphere.   The contrast of the active region is $C_{\rm s} = 1-({I_{\rm a}}/{I_{\rm phot}})$, 
where  $I_{\rm phot}$ is the brightness of the unperturbed photosphere.  For an active region dominated by the cool spot, $C_{\rm s} > 0$ because $I_{\rm a} < I_{\rm phot}$, while for an active region dominated by the hot spot,  $C_{\rm s} < 0$.
 
\begin{figure}
\centerline{\includegraphics[width=0.60\linewidth,angle=90]{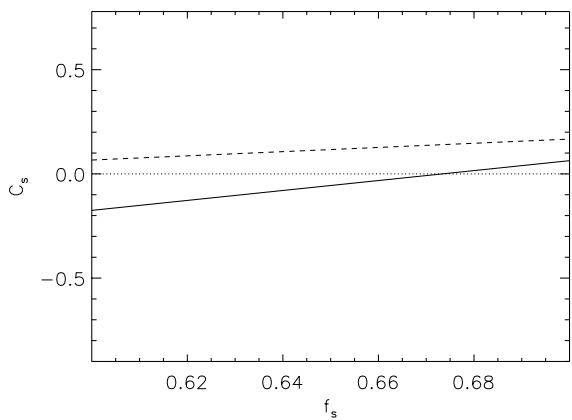}}
\caption{Contrast of an active region consisting of co-spatial cool and hot spots versus the fraction of its area covered by the cool spot. Two wavelengths are considered: $\lambda_{\rm opt} = 636$~nm (solid line) and $\lambda_{\rm NIR} = 1705$~nm (dashed line). }
\label{contrast}
\end{figure}

For BD+20~1790, we assume $T_{\rm phot} = 4410$~K (HO15), while for the cool and the hot spots  we assume temperatures $T_{\rm c} = T_{\rm phot}-1000$~K and $T_{\rm h} = T_{\rm phot}+1000$~K, respectively. Those temperature differences are typical of young and active stars such as Weak T-Tauri stars \citep[cf.][]{Rice,Koen16}. The contrast is plotted versus the fraction of the active region covered by the cool spot in Fig.~\ref{contrast} for the optical wavelength $\lambda_{\rm opt} = 636$~nm and the NIR wavelength $\lambda_{\rm NIR} = 1705$~nm. Those values correspond approximately to the mean wavelengths of the $gri$ bands and of the $JHK$ bands, respectively. 

\begin{figure}
\centerline{\includegraphics[width=0.60\linewidth,angle=0]{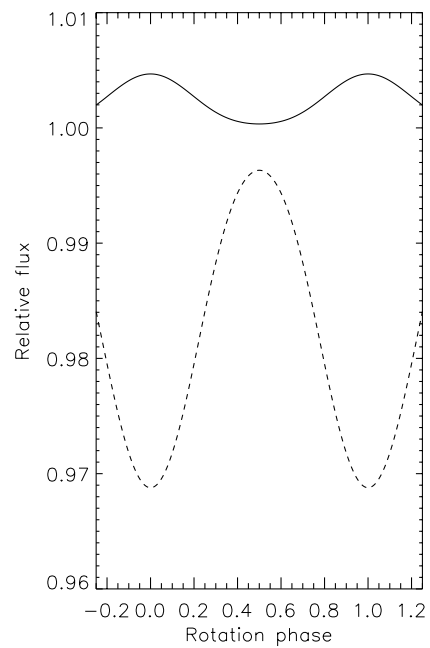}}
\caption{Light curves showing the rotational modulation of the flux at two wavelengths $\lambda_{\rm opt} = 636$~nm (solid line) and $\lambda_{\rm NIR} = 1705$~nm (dashed line). The amplitudes correspond to a spot having an area of 0.2 of the stellar disc. }
\label{light_curve}
\end{figure}

We see that for a restricted range of $f_{\rm s}$, that is $0.65 < f_{\rm s} < 0.67$, the contrast in the optical is negative and small, while that in the NIR is positive and remarkably larger. This leads to a rotational modulation of the optical flux remarkably smaller than, and in anti-phase with, the rotational modulation in the NIR as illustrated by the synthetic light curves in Fig.~\ref{light_curve} computed with the model in Sect.~3.2 of \citet{Lanza2016}.  Specifically, these light curves were computed for  an inclination of the spin axis to the line of sight $i=50^{\circ}$, considering an active region of an area of 20\% of that of the star's disc, centred at a latitude of $60^{\circ}$. The quadratic limb-darkening coefficients at the two wavelengths were taken from \citet{Claretetal12}, while $C_{\rm s} (\lambda_{\rm opt}) = -0.02$ and $C_{\rm s} (\lambda_{\rm NIR}) = 0.15$ corresponding to $f_{\rm s} \simeq 0.665$ in our simple irradiance model (cf. Fig.~\ref{contrast}). For simplicity, we assumed that the spot contrasts did not depend on the position on the disc.

The present model is simply illustrative. The amplitudes of the synthesized light curves are remarkably smaller than those observed in BD+20~1790, which may require a larger filling factor of the active region and/or  larger temperature contrasts. For example, in the case of LkCa~4, \citet{GullySantiagoetal17} found a filling factor as large as 86\%, which would imply an amplitude larger by a factor of approximately four in the case of the present model. 
We finally notice that the variable characteristics of the activity of this star with time might explain why past studies obtained quite different values for the r.m.s. scatter of the RVs, and then different interpretations of the nature of this object. 

As previously mentioned, the small amplitude of the RV modulation in the NIR as measured by GIANO contemporaneously with REM photometry is intriguing. Assuming that the depth of the spectral lines relative to their adjacent continuum is constant and considering a spot with a contrast $C_{\rm s}$ at latitude $\phi$ with a filling factor $f_{\rm s}$, we expect an RV modulation approximately of $ C_{\rm s} f_{\rm s} v\sin i \cos \phi$ (cf. \citealt{Saar1997}; \citealt{Desort2007}) that is  $\approx 0.75-1.0$ km~s$^{-1}$ for a spot at $\phi = 60^{\circ}$ , which is remarkably higher than what has been observed. A cool spot at a higher latitude would reduce the amplitude of the NIR wide-band flux modulations; a quenching of the convective shifts or the Zeeman effect also do not appear to be viable explanations because they increase the effect of a cool spot on the RV at NIR wavelengths \citep{Reiners2013}. Nevertheless, the variation of the relative depths of the spectral lines in the NIR cannot be neglected and it is the dominant effect in the cool spot responsible for the large photometric modulation in the infrared. The relative line depths are strong functions of the continuum opacity and of the degree of element ionization both remarkably varying in a cool spot area with respect to the unperturbed photosphere. In general, these effects produce a remarkable increase of the relative depths of the spectral lines in the cool spot. This  compensates for the decrease of the continuum intensity in the spot,  reducing the distortions of the spectral line profiles and yielding an RV variation in the NIR smaller than expected from the wide-band photometric variation in the case of constant relative line depths. In any case, a quantitative analysis is not warranted by our data, since a larger number of observations would be required.
This scenario suggests a need to better investigate this kind of target, since they might go through specific activity phases during which the VIS and NIR RV amplitudes are similar, possibly resulting in false positives. Looking at the curve phase shifts might give crucial information in these cases (see e.g. the recent result by \citealt{hatzes} for the K-giant $\gamma$ Draconis).


\section{Conclusions}
\label{sec:conc}
In this paper we present the analysis of RV measurements of the star BD+20~1790, in order to resolve the debate on the presence of a hot Jupiter with an orbital period of 7.8 days, claimed by HO10 and questioned by \cite{Figueira}. Since all the previous RV measurements were provided by visible spectrographs, we observed this target with the GIANO (in a dedicated run and during the GIANO -- B/GIARPS Commissioning) and IGRINS NIR spectrographs to disentangle the origin of the RV variation. The NIR data show a peak-to-valley amplitude of 437.3 ms$^{-1}$, significantly lower than the VIS one (1853.6 ms$^{-1}$), demonstrating that the detected RV modulation is due to activity. Therefore, we dismiss the presence of the claimed hot Jupiter around BD+20~1790, which is a very peculiar target from the point of view of stellar activity, as confirmed by our photometric monitoring with REM almost contemporary to the first spectroscopic run with GIANO. Surprisingly, REM light curves show larger amplitudes at longer wavelengths and the NIR curve is almost in anti-phase with respect to the visible one. This has been explained with a photometric model that includes a mixture of cool and hot spots in the same active region.\\
\indent Multi-band spectroscopy is known to be crucial in the search for exoplanets around young and active stars: here we present the first contribution of a new facility, GIARPS, which adds further value to this method thanks to the simultaneous observations in VIS and NIR bands. Our result clears the current census of hot Jupiters from the only previously known case orbiting a star between 20 to 200 Myr old \citep{Rizzuto2017, DavidT2018}. The paucity of hot Jupiters in this age range might be explained by the time dependence of planet migration mechanisms (Mann et al. 2018 \footnote{\url{https://keplerscience.arc.nasa.gov/cluster-workshop/talks/k2clusters-13-andrew-mann.pdf}.}) or simply by small number statistics. Further investigations are needed to achieve a firm conclusion on this key issue for the evolution of planetary systems.

\begin{acknowledgements}
The authors are grateful to the anonymous referee for the careful review and the useful comments that helped to improve the quality of the paper. The authors acknowledge support from INAF through the "WOW Premiale" funding scheme of the Italian Ministry of Education, University, and Research. 
\end{acknowledgements}

\longtab{
\begin{longtable}{llcccc}
\caption{\label{tab:bd201790_giano} Time series of BD+20~1790 from GIANO and GIANO -- B data. For each observation we list radial velocities (RV) and the bisector span (BIS) with the corresponding uncertainties.}\\
\hline
\noalign{\smallskip}
Dataset & JD-2450000  &      RV         &    $\sigma_{RV}$     &   $BIS$     & $\sigma_{BIS}$  \\
        &     & (km\, s$^{-1}$) &    (km\, s$^{-1}$)   &  (km\, s$^{-1}$)   &  (km\, s$^{-1}$)  \\
\hline
\noalign{\smallskip}
GIANO        &  7475.8302179 &  8.476         &    0.045                  &   0.090             &   0.071\\
             &  7475.8387828 &  8.539         &    0.045                  &   0.138             &   0.074\\
             &  7475.8473007 &  8.552         &    0.045                      &   0.082         &   0.076\\
             &  7476.8292450 &  8.626         &    0.045                  &        0.289        &   0.076\\
             &  7476.8378800 &  8.637         &    0.045                  &        0.106        &   0.069\\
             &  7476.8463629 &  8.584         &     0.045                 &        0.232        &   0.073\\
             &  7478.8578259 &  8.616         &     0.045                 &        0.043        &   0.083\\
             &  7478.8768308 &  8.626         &     0.045                 &        0.128        &   0.083\\
             &  7478.8946776 &  8.556         &     0.045                 &        0.036        &   0.078\\
             &  7478.9032996 &  8.602         &     0.045                 &        0.124        &   0.076\\
             &  7479.8734209 &  8.523         &     0.045                 &        0.123        &   0.049\\
             &  7479.8819158 &  8.586         &     0.045                 &        0.017        &   0.052\\
             &  7479.8917888 &  8.566         &     0.045                 &        0.106        &   0.054\\
             &  7480.8530107 &  8.588         &     0.045                 &        0.226        &   0.076\\
             &  7480.8612976 &  8.550         &     0.045                 &        0.147        &   0.072\\
             &  7481.8545340 &  8.560         &     0.045                 &        -0.005           &   0.071\\
             &  7481.8631110 &  8.491         &     0.045                 &        0.178        &   0.070\\
             &  7481.8715830 &  8.550         &     0.045                 &        0.119        &   0.067\\
\hline
GIANO -- B      &  7716.7390467 &       8.256          &    0.022                 &        0.183        &   0.026\\
             &  7716.7477729 &  8.263          &    0.022                     &    0.177        &   0.026\\
             &  7716.7564760 &  8.253          &    0.022                     &    0.208        &   0.025\\
             &  7717.5770302 &  8.256          &    0.022                     &    0.125        &   0.039\\
             &  7717.5857217 &  8.335          &    0.022                     &    0.141        &   0.036\\
             &  7717.5945405 &  8.266          &    0.022                     &    0.170        &   0.031\\
             &  7717.6046440 &  8.273          &    0.022                     &    -0.016       &   0.036\\
\hline
GIANO -- B      &  7821.4345852 &       8.474          &    0.036                     &   -0.007        &   0.085\\
             &  7821.4436589 &  8.498          &    0.036                     &    -0.014           &   0.084\\
             &  7827.4492982 &  8.440          &    0.036                     &    0.332        &   0.087\\
             &  7827.4575966 &  8.460          &    0.036                     &    0.274        &   0.090\\
             &  7827.4658833 &  8.437          &    0.036                     &    0.094        &   0.092\\
\hline
\noalign{\smallskip}
\end{longtable}
}
\vspace{1cm}
\longtab{
\begin{longtable}{lcccccc}
\caption{\label{tab:bd201790_harpsn} Time series of BD+20~1790 from HARPS -- N data, from HO15 and GIARPS commissioning, uniformly reduced with the new HARPS -- N DRS version. For each observation we list radial velocities (RV), $\log R'_{HK}$ , and the bisector span (BIS) with their related uncertainties.}\\
\hline
\noalign{\smallskip}
 JD-2450000  &      RV         &    $\sigma_{RV}$      & $BIS$    & $\sigma_{BIS}$ &  $\log R'_{HK}$ & $\sigma_{\log R'_{HK}}$  \\
             & (km\, s$^{-1}$) &    (km\, s$^{-1}$)   &   (km\, s$^{-1}$)   &  (km\, s$^{-1}$) &  & \\
\hline             
\noalign{\smallskip}
6293.496128  &  7.818  &  0.004  &    0.019  &  0.009 &   -3.982  &  0.002\\
6293.694395  &  7.750  &  0.004  &    0.065  &  0.021 &   -3.984  &  0.003 \\
6293.768482  &  7.726  &  0.005  &    0.120  &  0.020 &   -3.995  &  0.003\\
6346.563090  &  8.021  &  0.003  &   -0.247  &  0.009 &   -3.944  &  0.002 \\
6347.423110  &  7.374  &  0.007  &    0.547  &  0.017 &   -3.977  &  0.004\\
6347.601234  &  7.163  &  0.005  &    0.754  &  0.017 &   -3.989  &  0.004 \\
6348.543400  &  8.113  &  0.008  &   -0.217  &  0.021 &   -4.016  &  0.007  \\
6348.548909  &  8.113  &  0.010  &   -0.207  &  0.024 &   -4.041  &  0.008  \\
6348.620652  &  8.167  &  0.009  &   -0.290  &  0.022 &   -4.018  &  0.008 \\
6348.626160  &  8.166  &  0.008  &   -0.242  &  0.020 &   -4.022  &  0.007  \\
6349.437201  &  8.017  &  0.015  &   -0.174  &  0.036 &   -3.946  &  0.011 \\
6349.442721  &  8.027  &  0.011  &   -0.225  &  0.028 &   -3.955  &  0.008\\
6679.499824  &  8.346  &  0.004  &   -0.493  &  0.013 &   -3.972  &  0.003 \\
6679.680352  &  8.177  &  0.004  &   -0.409  &  0.018 &   -3.887  &  0.002 \\
6682.562820  &  8.137  &  0.015  &   -0.322  &  0.038 &   -3.941  &  0.011 \\
6682.666891  &  7.945  &  0.009  &   -0.124  &  0.026 &   -3.966  &  0.007  \\
6796.370888  &  7.748  &  0.003  &    0.013  &  0.012 &   -4.030  &  0.003 \\
7821.424255  &  7.690  &  0.013  &    0.097  &  0.033 &   -4.023  &  0.009\\
7823.488210  &  8.031  &  0.013  &   -0.250  &  0.041 &   -3.946  &  0.009 \\
7827.465395  &  7.915  &  0.004  &    0.015  &  0.008 &   -3.991  &  0.003 \\
\hline
\noalign{\smallskip}
\end{longtable}
}
\vspace{1cm}
\longtab{
\begin{longtable}{llcccc}
\caption{\label{tab:bd201790_igrins} Time series of BD+20~1790 from IGRINS data. For each observation we list radial velocities (RV) and the bisector span (BIS) with the corresponding uncertainties.}\\
\hline
\noalign{\smallskip}
Dataset     &JD-2450000  &      RV         &    $\sigma_{RV}$     &   $BIS$     & $\sigma_{BIS}$  \\
        &     & (km\, s$^{-1}$) &    (km\, s$^{-1}$)   &  (km\, s$^{-1}$)   &  (km\, s$^{-1}$)  \\
\hline
\noalign{\smallskip}
IGRINS@McD & 7417.80148148 &     8.106         &        0.059             &   -0.249            &    0.062\\
            & 7418.77283564 &    8.220         &        0.059             &    0.025            &    0.029\\
            & 7419.81490740 &    8.146         &        0.059             &   -0.420            &    0.065\\
            & 7421.84695601 &    8.037         &        0.059             &   -0.127            &    0.075\\
            & 7475.63379629 &    8.133         &        0.059             &    0.438            &    0.065\\
            & 7476.61835648 &    8.161         &        0.059             &    0.221            &    0.062\\
            & 7503.59172453 &  8.205         &  0.059             &   -0.030           &    0.051\\
\hline
IGRINS@DCT  & 7671.96244213 &  8.285         &  0.065             &    0.362           &    0.054\\
            & 7673.95834490 &  8.015         &  0.065             &    0.270           &    0.056\\
            & 7675.86873842 &  8.178         &  0.065             &    0.339           &    0.047 \\
            & 7713.04495370 &  8.219         &    0.065             &    0.143           &    0.060\\
            & 7732.03075231 &  8.248         &  0.065             &    0.117           &    0.054\\
            & 7742.97497685 &  8.244         &  0.065             &    0.464           &    0.078\\
            & 7793.85260416 &  8.021         &  0.065             &    0.228           &    0.043\\
            & 7794.84053240 &  8.035         &  0.065             &    0.345           &    0.033\\
            & 7796.74128472 &  8.041         &  0.065             &    0.181           &    0.072\\
            & 7798.78775463 &  7.953         &  0.065             &    0.396           &    0.047\\
\hline
IGRINS@McD  & 7825.74646990 &  8.018         &  0.044             &    0.318           &    0.063\\
            & 7827.67596064 &  8.131         &  0.044             &   -0.125           &    0.053\\
            & 7828.69434027 &  8.046         &  0.044             &    0.136           &    0.056\\
            & 7829.75975694 &  7.958         &  0.044             &   -0.053           &    0.071\\
            & 7862.63905092 &  7.964         &  0.044             &   -0.101           &    0.053\\
            & 7864.69674768 &  7.973         &  0.044             &    0.144           &    0.055\\
            & 7875.58734953 &  7.925         &  0.044             &    0.026           &    0.053\\
            & 7876.59074074 &  8.079         &  0.044             &    0.213           &    0.055\\
            & 7877.58625000 &  7.902         &  0.044             &    0.134           &    0.055\\
            & 7878.58519676 &  7.986         &  0.044             &    0.179           &    0.058\\
            & 7879.58496527 &  8.017         &  0.044             &   -0.006           &    0.053\\
            & 7880.58305555 &  7.965         &  0.044             &   -0.023           &    0.052\\
\hline
\noalign{\smallskip}
\end{longtable}
}
\vspace{1cm}

\longtab{
\begin{longtable}{cccc}
\caption{\label{tab:rem} Time series of BD+20~1790 from REM data with different filters. For each observation we list the differential magnitudes with the corresponding uncertainty.}\\
\hline
\noalign{\smallskip}
Filter   &  JD-2450000     &  $\Delta m$     &   $\sigma_{\Delta m}$  \\
         &                 &  (mag)     &   (mag)              \\
\hline  
\noalign{\smallskip}
g        &  7477.52744     & -1.960                  &         0.005               \\
         &  7478.58315     & -1.985                  &         0.004               \\   
         &  7479.57717     & -2.002                  &         0.007               \\   
         &  7480.59303     & -1.984                  &         0.008               \\   
         &  7481.59388     & -2.012                  &         0.009               \\   
         &  7482.61009     & -2.010                  &         0.008               \\   

\noalign{\smallskip}
\hline
\noalign{\smallskip}
r        &  7477.52744     & -2.687                  &         0.013              \\
         &  7478.58315     & -2.722                  &         0.017               \\
         &  7479.58383     & -2.732                  &         0.016               \\
         &  7480.59303     & -2.725                  &         0.012               \\
         &  7481.59283     & -2.753                  &         0.011               \\
         &  7482.61159     & -2.756                  &         0.013               \\
\noalign{\smallskip}
\hline
\noalign{\smallskip}
 i       &  7477.52822     & -3.100                  &         0.007              \\  
         &  7478.58392     & -3.109                  &         0.011               \\
         &  7479.57717     & -3.129                  &         0.007               \\
         &  7480.58715     & -3.104                  &         0.006               \\
         &  7481.58804     & -3.124                  &         0.008               \\
         &  7482.60679     & -3.122                  &         0.010               \\
\noalign{\smallskip}
\hline
\noalign{\smallskip}
z       &  7477.52744     & -3.318                  &         0.010               \\ 
        &  7478.58315     & -3.279                  &         0.011               \\ 
        &  7479.57717     & -3.344                  &         0.010               \\ 
        &  7480.59196     & -3.348                  &         0.011               \\ 
        &  7481.59097     & -3.320                  &         0.012               \\ 
        &  7482.61009     & -3.409                  &         0.015               \\ 

\noalign{\smallskip}
\hline
\noalign{\smallskip}
J       & 7479.57544     & -3.756                  &         0.006               \\
        & 7480.58707     & -3.839                  &         0.020               \\
        & 7481.58691     & -3.687                  &         0.005               \\
        & 7482.60688     & -3.756                  &         0.035               \\

\noalign{\smallskip}
\hline
\noalign{\smallskip}
H       &  7479.57789     & -3.527                  &         0.044               \\
        &  7480.58792     & -3.762                  &         0.043               \\
        &  7481.58772     & -3.575                  &         0.033               \\
        &  7482.60755     & -3.734                  &         0.034               \\
\noalign{\smallskip}
\hline
\noalign{\smallskip}
K       &  7479.58154     & -3.850                  &         0.067               \\  
        &  7480.59008     & -4.003                  &         0.014               \\  
        &  7481.58988     & -3.801                  &         0.040               \\  
        &  7482.60970     & -3.904                  &         0.044               \\  
\noalign{\smallskip}
\hline
\end{longtable}
}

\end{document}